\documentclass[prl,superscriptaddress,twocolumn]{revtex4}%
\usepackage[ansinew]{inputenc}
\usepackage[dvips]{graphicx}
\usepackage{amsmath}
\usepackage{amsthm}
\usepackage{bm}
\usepackage{layout}
\usepackage{float}
\usepackage{txfonts}
\usepackage{amsfonts}
\usepackage{amssymb}%
\begin{document}

\title{Classical world arising out of quantum physics under the restriction of
coarse-grained measurements}

\begin{abstract}
Conceptually different from the decoherence program, we present a novel
theoretical approach to macroscopic realism and classical physics within
quantum theory. It focuses on the limits of observability of quantum effects
of macroscopic objects, i.e., on the required precision of our measurement
apparatuses such that quantum phenomena can still be observed. First, we
demonstrate that for unrestricted measurement accuracy no classical
description is possible for arbitrarily large systems. Then we show for a
certain time evolution that under coarse-grained measurements not only
macrorealism but even the classical Newtonian laws emerge out of the
Schr\"odinger equation and the projection postulate.

\end{abstract}
\date{\today}%

\author{Johannes Kofler}%
%

\affiliation{Fakult\"at f\"ur Physik, Universit\"{a}%
t Wien, Boltzmanngasse 5, 1090 Wien, Austria}%
%

\affiliation
{Institut f\"ur Quantenoptik und Quanteninformation (IQOQI), \"Osterreichische Akademie der Wissenschaften,\\ Boltzmanngasse 3, 1090 Wien, Austria}%
%

\author{{\v C}aslav Brukner}%
%

\affiliation{Fakult\"at f\"ur Physik, Universit\"{a}%
t Wien, Boltzmanngasse 5, 1090 Wien, Austria}%
%

\affiliation
{Institut f\"ur Quantenoptik und Quanteninformation (IQOQI), \"Osterreichische Akademie der Wissenschaften,\\ Boltzmanngasse 3, 1090 Wien, Austria}%
%

\maketitle

Quantum physics is in conflict with a classical world view both conceptually
and mathematically. The assumptions of a genuine classical world---local
realism and macroscopic realism---are at variance with quantum mechanical
predictions as characterized by the violation of the Bell and Leggett--Garg
inequality, respectively~\cite{Bell1964,Legg1985}. Does this mean that the
classical world is substantially different from the quantum world? When and
how do physical systems stop to behave quantumly and begin to behave
classically? Although questions like these date back to Schrödinger's famous
cat paper~\cite{Schr1935}, the opinions in the physics community still differ
dramatically. Various views range from the mere experimental difficulty of
sufficiently isolating any system from its environment
(decoherence)~\cite{Zure1991} to the principal impossibility of superpositions
of macroscopically distinct states due to the breakdown of quantum physical
laws at some quantum-classical border (collapse models)~\cite{Ghir1986}.

Macrorealism is defined by the conjunction of two postulates~\cite{Legg1985}:
\textquotedblright\textit{Macrorealism per se}: A macroscopic object which has
available to it two or more macroscopically distinct states is at any given
time in a definite one of those states. \textit{Non-invasive measurability}:
It is possible in principle to determine which of these states the system is
in without any effect on the state itself or on the subsequent system
dynamics.\textquotedblright\ These assumptions allow to derive the
Leggett--Garg inequalities.

In this Letter---inspired by the thoughts of Peres on the classical
limit~\cite{Pere1995}---we present a novel theoretical approach to macroscopic
realism and classical physics \textit{within} quantum theory. We first show
that, if consecutive eigenvalues $m$ of a spin component can sufficiently be
experimentally resolved, a \textit{Leggett--Garg inequality will be violated
for arbitrary spin lengths} $j$ and the violation persists even in the limit
of infinitely large spins. This contradicts the naive assumption that the
predictions of quantum mechanics reduce to those of classical physics when a
system becomes \textquotedblright large\textquotedblright\ and was
demonstrated for local realism by Garg and Mermin~\cite{Garg1982}. Note that
due to the resolution of consecutive eigenvalues one cannot speak about
violation of macrorealism. If, however, for a certain time evolution one goes
into the limit of large spin lengths but can experimentally only resolve
eigenvalues $m$ which are separated by much more than the square root of the
spin length (the intrinsic quantum uncertainty), i.e., $\Delta m\gg\!\sqrt{j}%
$, the macroscopically distinct outcomes appear to obey classical (Newtonian)
laws. This suggests that \textit{macrorealism and classical laws emerge out of
quantum physics under the restriction of coarse-grained measurements}.

While our approach is not at variance with the decoherence program, it differs
conceptually from it. It is not dynamical and puts the stress on the limits of
observability of quantum effects of macroscopic objects. The term
\textquotedblright macroscopic\textquotedblright\ throughout the paper is used
to denote a system with a high dimensionality rather than a low-dimensional
system with a large parameter such as mass or size.

Consider a physical system and a quantity $Q$, which whenever measured is
found to take one of the values $\pm1$ only. Further consider a series of runs
starting from identical initial conditions such that on the first set of runs
$Q$ is measured only at times $t_{1}$ and $t_{2}$, only at $t_{2}$ and $t_{3}$
on the second, at $t_{3}$ and $t_{4}$ on the third, and at $t_{1}$ and $t_{4}$
on the fourth $(0\leq t_{1}<t_{2}<t_{3}<t_{4})$. Introducing temporal
correlations $C_{ij}\equiv\langle Q(t_{i})\,Q(t_{j})\rangle$, any
macrorealistic theory predicts the Leggett--Garg inequality~\cite{Legg1985}%
\begin{equation}
K\equiv C_{12}+C_{23}+C_{34}-C_{14}\leq2\,. \label{eq Leggett}%
\end{equation}
This inequality is violated, e.g., by the precession of a spin-$\frac{1}{2}$
particle with the Hamiltonian $\hat{H}=\frac{1}{2}\,\omega\,\hat{\sigma}_{x}$
with $\omega$ the angular precession frequency and $\hat{\sigma}_{x}$ the
Pauli $x$-matrix. (We use units in which the reduced Planck constant is
$\hbar=1$). Measuring the spin along the $z$-direction, we obtain the
correlations $C_{ij}=\langle\hat{\sigma}_{z}(t_{i})\,\hat{\sigma}_{z}%
(t_{j})\rangle=\cos[\omega(t_{j}\!-\!t_{i})]$. Choosing, e.g., equidistant
measurement times with time difference $\Delta t=\pi/4\omega$,
ineq.~(\ref{eq Leggett}) is violated as $K=2\sqrt{2}$, which is understandable
since a spin-$\frac{1}{2}$ particle is a genuine quantum object. In contrast,
any rotating classical spin vector always satisfies the inequality.

In the following, we show that the Leggett--Garg inequality (\ref{eq Leggett})
is violated for arbitrarily large spin lengths $j$. As the first measurement
will act as a preparation of the state for the subsequent measurement, the
initial state is not decisive and it is sufficient to consider the maximally
mixed state%
\begin{equation}
\hat{\rho}(0)\equiv\dfrac{1}{2j+1}\,%
{\displaystyle\sum\nolimits_{m=-j}^{j}}
\left\vert m\right\rangle \!\left\langle m\right\vert =\dfrac{\openone}{2j+1}
\label{eq mixed state}%
\end{equation}
with $\openone$ the identity operator and $\left\vert m\right\rangle $ the
$\hat{J}_{z}$ (spin $z$-component) eigenstates. The Hamiltonian be%
\begin{equation}
\hat{H}=\hat{\mathbf{J}}^{2}/2I+\omega\,\hat{J}_{x}\,, \label{eq Hamiltonian}%
\end{equation}
where $\hat{\mathbf{J}}$ is the rotor's total spin vector operator, $\hat
{J}_{x}$ its $x$-component, $I$ the moment of inertia and $\omega$ the angular
precession frequency. Here, $\hat{\mathbf{J}}^{2}/2I$ commutes with the
individual spin components and does not contribute to the time evolution. The
solution of the Schrödinger equation produces a rotation about the $x$-axis,
represented by the time evolution operator $\hat{U}_{t}=\;$e$^{-\text{i}\omega
t\hat{J}_{x}}$ with $t$ the time. We define the parity measurement $\hat
{Q}\equiv%
{\textstyle\sum\nolimits_{m=-j}^{j}}
(-1)^{j-m}\left\vert m\right\rangle \!\left\langle m\right\vert =$%
e$^{\text{i}\pi(j-\hat{J}_{z})}$ with possible dichotomic outcomes $\pm$
(identifying $\pm\equiv\pm1$). The correlation function between results of the
parity measurement $\hat{Q}$ at different times $t_{1}$ and $t_{2}$ is
$C_{12}\equiv p_{+}\,q_{+|+}+p_{-}\,q_{-|-}-p_{+}\,q_{-|+}-p_{-}\,q_{+|-}$,
where $p_{+}$ ($p_{-}$) is the probability for measuring $+$ ($-$) at $t_{1}$
and $q_{l|k}$ is the probability for measuring $l$ at $t_{2}$ given that $k$
was measured at $t_{1}$ ($k,l=+,-$). Furthermore, $p_{+}=1-p_{-}=\frac{1}%
{2}\,(\langle\hat{Q}_{t_{1}}\rangle+1)$, $q_{+|\pm}=1-q_{-|\pm}=\frac{1}%
{2}\,(\langle\hat{Q}_{t_{2}}\rangle_{\pm}+1)$. Here $\langle\hat{Q}_{t_{1}%
}\rangle$ is the expectation value of $\hat{Q}$ at $t_{1}$ and $\langle\hat
{Q}_{t_{2}}\rangle_{\pm}$ is the expectation value of $\hat{Q}$ at $t_{2}$
given the outcome $\pm$ at $t_{1}$.

Using $\hat{\rho}(t_{1})=\hat{U}_{t_{1}}\,\hat{\rho}(0)\,\hat{U}_{t_{1}%
}^{\dagger}=\hat{\rho}(0)$, we find $\langle\hat{Q}_{t_{1}}\rangle\equiv
\;$Tr$[\hat{\rho}(t_{1})\,\hat{Q}]\approx0$. The approximate sign is accurate
for half integer $j$ and in the macroscopic limit $j\gg1$, which is assumed
from now on. Hence, as expected we have $p_{+}=\tfrac{1}{2}$. Depending on the
measurement result at $t_{1}$, the state is reduced to $\hat{\rho}_{\pm}%
(t_{1})\equiv\hat{P}_{\pm}\,\hat{\rho}(t_{1})\,\hat{P}_{\pm}/$Tr$[\hat{P}%
_{\pm}\,\hat{\rho}(t_{1})\,\hat{P}_{\pm}]=(\openone\pm\hat{Q})/(2j+1)$ with
$\hat{P}_{\pm}\equiv\tfrac{1}{2}(\openone\pm\hat{Q})$ the projection operator
onto positive (negative) parity states. Denoting $\Delta t\equiv t_{2}-t_{1}$
and $\theta\equiv\omega\,\Delta t$, we obtain $\langle\hat{Q}_{t_{2}}%
\rangle_{\pm}\equiv\;$Tr$[\hat{U}_{\Delta t}\,\hat{\rho}_{\pm}(t_{1})\,\hat
{U}_{\Delta t}^{\dagger}\,\hat{Q}]=\pm$Tr$[$e$^{2\text{i}\theta\hat{J}_{x}%
}]/(2j+1)=\pm\sin[(2j+1)\,\omega\,\Delta t]/(2j+1)\sin[\omega\,\Delta t]$.

From $\langle\hat{Q}_{t_{2}}\rangle_{+}=-\langle\hat{Q}_{t_{2}}\rangle_{-}$ it
follows $q_{+|+}+q_{+|-}=1$. Using this and $p_{+}=\tfrac{1}{2}$, the temporal
correlation becomes $C_{12}=\left\langle \!\right.  \hat{Q}_{t_{2}}\left.
\!\right\rangle _{+}$. With equidistant times, time distance $\Delta t$, and
the abbreviation $x\equiv(2j+1)\,\omega\,\Delta t$ the Leggett--Garg
inequality (\ref{eq Leggett}) reads%
\begin{equation}
K\approx\dfrac{3\sin x}{x}-\dfrac{\sin3x}{3x}\leq2\,. \label{eq Leggett K(x)}%
\end{equation}
The sine function in the denominator was approximated, assuming $\tfrac
{x}{2j+1}\ll1$. Inequality~(\ref{eq Leggett K(x)}) is violated for all
positive $x\lesssim1.656$ and maximally violated for $x\approx1.054$ where
$K\approx2.481$ (compare with Ref.~\cite{Pere1995} for the violation of local
realism). \textit{We can conclude that a violation of the Leggett--Garg
inequality is possible for arbitrarily high-dimensional systems and also for
totally mixed states}, given that consecutive values of $m$ can be resolved.

In the second part of the paper we will show that inaccurate measurements not
only lead to validity of macrorealism but even to the \textit{emergence of
classical physics}.

In quantum theory any two different eigenvalues $m_{1}$ and $m_{2}$ in a
measurement of a spin's $z$-component correspond to orthogonal states
\textit{without any concept of closeness or distance}. The terms
\textquotedblright close\textquotedblright\ or \textquotedblright
distant\textquotedblright\ only make sense in a classical context, where those
eigenvalues are treated as close which correspond to \textit{neighboring}
outcomes in the real configuration space. For example, the \textquotedblright
eigenvalue labels\textquotedblright\ $m$ and $m\!+\!1$ of a spin observable
correspond to neighboring outcomes in a Stern-Gerlach experiment. (Such
observables are sometimes called classical or
reasonable~\cite{Yaff1982,Pere1995}.) It is those neighboring eigenvalues
which we conflate to coarse-grained observables in measurements of limited
accuracy. It seems thus unavoidable that certain features of classicality have
to be assumed beforehand.

In what follows we will first consider the special case of a single spin
coherent state and then generalize the transition to classicality for
arbitrary states. \textit{Spin-}$j$\textit{ coherent states }$\left\vert
\vartheta,\varphi\right\rangle $~\cite{Radc1971} are the eigenstates with
maximal eigenvalue of a spin operator pointing into the ($\vartheta,\varphi
$)-direction, where $\vartheta$ and $\varphi$ are the polar and azimuthal
angle, respectively: $\hat{\mathbf{J}}\left\vert \vartheta,\varphi
\right\rangle =j\left\vert \vartheta,\varphi\right\rangle $. At time $t=0$ let
us consider $|\vartheta_{0},\varphi_{0}\rangle=%
{\textstyle\sum\nolimits_{m}}
\left(  \!%
\genfrac{}{}{0pt}{1}{2j}{j+m}%
\!\right)  \!^{1/2}\cos^{j+m}\!\tfrac{\vartheta_{0}}{2}\,\sin^{j-m}%
\!\tfrac{\vartheta_{0}}{2}\;$e$^{-\text{i}m\varphi_{0}}\,|m\rangle$. Under
time evolution $\hat{U}_{t}=\;$e$^{-\text{i}\omega t\hat{J}_{x}}$ the
probability that a $\hat{J}_{z}$ measurement at time $t$ has outcome $m$ is
$p(m,t)=|\langle m|\vartheta,\varphi\rangle|^{2}$ with $\cos\vartheta
=\sin\omega t\,\sin\vartheta_{0}\,\sin\varphi_{0}+\cos\omega t\,\cos
\vartheta_{0}$, where $\vartheta$ and $\varphi$ are the polar and azimuthal
angle of the (rotated) spin coherent state $\left\vert \vartheta
,\varphi\right\rangle $ at time $t$. In the macroscopic limit, $j\gg1$, the
binomial can be well approximated by a Gaussian distribution%
\begin{equation}
p(m,t)\approx\dfrac{1}{\sqrt{2\pi}\,\sigma}\,\text{e}^{-(m-\mu)^{2}%
/2\sigma^{2}} \label{eq p(m1,t1)}%
\end{equation}
with $\sigma\equiv\sqrt{j/2}\,\sin\vartheta$ the width and $\mu\equiv
j\cos\vartheta$ the mean.

\begin{figure}[t]
\begin{center}
\includegraphics{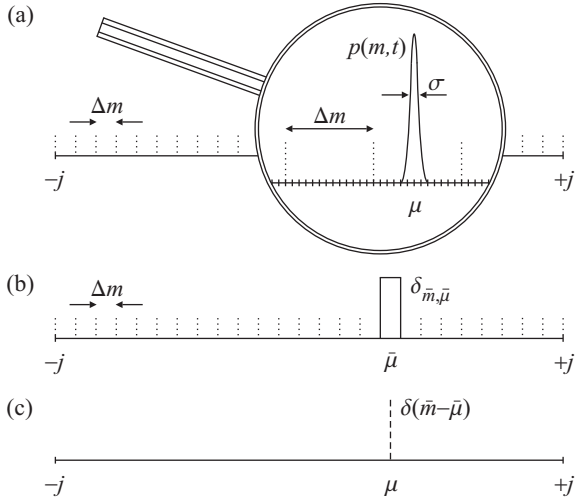}
\end{center}
\par
\vspace{-0.357cm}\caption{An initial spin-$j$ coherent state $\left\vert
\vartheta_{0},\varphi_{0}\right\rangle $ precesses into the coherent state
$\left\vert \vartheta,\varphi\right\rangle $ at time $t$ under a quantum time
evolution. (a)~The probability $p(m,t)$ for the outcome $m$ in a measurement
of the spin's $z$-component is given by a Gaussian distribution with width
$\sigma$ and mean $\mu$, which can be seen under the magnifying glass of sharp
measurements. (b)~The measurement resolution $\Delta m$ is finite and
subdivides the $2j+1$ possible outcomes into a smaller number of
coarse-grained \textquotedblright slots\textquotedblright. If the measurement
accuracy is much poorer than the width $\sigma$, i.e., $\Delta m\gg\!\sqrt{j}%
$, the sharply peaked Gaussian cannot be distinguished anymore from the
discrete Kronecker delta $\delta_{\bar{m},\bar{\mu}}$ where $\bar{m}$ is
numbering the slots and $\bar{\mu}$ is the slot in which the center $\mu$ of
the Gaussian lies. (c)~In the limit $j\rightarrow\infty$ the slots
\textit{seem} to become infinitely narrow and $\delta_{\bar{m},\bar{\mu}}$
becomes the Dirac delta function $\delta(\bar{m}\!-\!\bar{\mu})$.}%
\label{fig1}%
\end{figure}

Under the \textquotedblright magnifying glass\textquotedblright\ of sharp
measurements we can see separate eigenvalues $m$ and resolve the Gaussian
probability distribution $p(m,t)$, as shown in Fig.~\ref{fig1}(a). Let us now
assume that the \textit{resolution of the measurement apparatus}, $\Delta m$,
is finite and subdivides the $2j+1$ possible outcomes $m$ into a smaller
number of $\frac{2j+1}{\Delta m}$ coarse-grained \textquotedblright
slots\textquotedblright. If the slot size is much larger than the standard
deviation $\sigma\sim\!\sqrt{j}$, i.e., $\Delta m\gg\!\sqrt{j}$, the sharply
peaked Gaussian cannot be distinguished anymore from the discrete Kronecker
delta,%
\begin{equation}
\Delta m\gg\!\sqrt{j}:\quad p(m,t)\rightarrow\delta_{\bar{m},\bar{\mu}}\,,
\label{eq Kronecker}%
\end{equation}
where $\bar{m}$ is numbering the slots (from $-j+\frac{\Delta m}{2}$ to
$j-\frac{\Delta m}{2}$ in steps $\Delta m$) and $\bar{\mu}$ is the number of
the slot in which the center $\mu$ of the Gaussian lies, as indicated in
Fig.~\ref{fig1}(b). In the \textit{limit of infinite dimensionality},
$j\rightarrow\infty$, one can distinguish two cases: (1) If the inaccuracy
$\Delta m$ scales linearly with $j$, i.e, $\Delta m=O(j)$, the discreteness
remains. (2) If $\Delta m$ scales slower than $j$, i.e., $\Delta m=o(j)$ but
still $\Delta m\gg\!\sqrt{j}$, then the slots \textit{seem} to become
infinitely \textit{narrow}. Pictorially, the \textit{real space length} of the
eigenvalue axis, representing the $2j+1$ possible outcomes $m$, is
\textit{limited} in any laboratory, e.g., by the size of the observation
screen after a Stern--Gerlach magnet, whereas the number of slots grows with
$j/\Delta m$. Then, in the limit $j\rightarrow\infty$, the Kronecker delta
becomes the Dirac delta function,%
\begin{equation}
\Delta m\gg\!\sqrt{j}\;\;\&\;\;j\rightarrow\infty:\quad p(m,t)\rightarrow
\delta(\bar{m}\!-\!\bar{\mu})\,, \label{eq delta}%
\end{equation}
which is shown in Fig.~\ref{fig1}(c).

Under a fuzzy measurement the reduced (projected) state is essentially the
state before the measurement. If $\left\vert \vartheta,\varphi\right\rangle $
is centered well inside the slot, the disturbance is exponentially small. Only
in the cases where it is close to the border between two slots, the
measurement is invasive. Assuming that the measurement times and/or slot
positions chosen by the observer are statistically independent of the
(initial) position of the coherent state, a disturbance happens merely in the
fraction $\sigma/\Delta m\ll1$ of all measurements. This is equivalent to the
already assumed condition $\!\sqrt{j}\ll\Delta m$. Therefore, fuzzy
measurements of a spin coherent state are largely non-invasive such as in any
macrorealistic theory, in particular classical Newtonian physics. Small errors
may accumulate over many measurements and eventually there might appear
deviations from the classical time evolution. This, however, is unavoidable in
any explanation of classicality \textit{gradually} emerging out of quantum
theory. To which extent this effect is relevant for our every-day experience
is an open issue~\footnote{For the general trade-off between measurement
accuracy and state disturbance for more realistic smoothed
positive-operator-valued measurements and for related approaches to
classicality see Refs.~\cite{Pere1995,Poul2005}.}.

Hence, at the coarse-grained level the physics of the (quantum) spin system
can completely be described by a \textquotedblright new\textquotedblright%
\ formalism, utilizing a (classical) \textit{spin vector} $\mathbf{J}$ at time
$t=0$, pointing in the ($\vartheta_{0},\varphi_{0}$)-direction with length
$J\equiv|\mathbf{J}|=\!\sqrt{j(j\!+\!1)}\approx j$, where $j\gg1$, and a
(Hamilton) function%
\begin{equation}
H=\mathbf{J}^{2}/2I+\omega\,J_{x}\,. \label{eq Hamilton}%
\end{equation}
At any time the probability that the spin vector's $z$-component
$J\cos\vartheta\approx j\cos\vartheta$ is in slot $\bar{m}$ is given by
$\delta_{\bar{m},\bar{\mu}}$, eq.~(\ref{eq Kronecker}), \textit{as if} the
time evolution of the spin components $J_{i}$ ($i=x,y,z$) is given by the
Poisson brackets, $\dot{J}_{i}=[J_{i},H]_{\text{PB}}$, and measurements are
non-invasive. Only the term $\omega\,J_{x}$ in eq.~(\ref{eq Hamilton}) governs
the time evolution and the solutions correspond to a rotation around the
$x$-axis. In the proper continuum limit the spin vector at time $t$ points in
the ($\vartheta,\varphi$)-direction where $\vartheta$ and $\varphi$ are the
same as for the spin coherent state and the prediction is given by
$\delta(\bar{m}\!-\!\bar{\mu})$, eq.~(\ref{eq delta}). \textit{This is
classical (Newtonian) mechanics}.

Finally, we show that the time evolution of \textit{any} spin-$j$ quantum
state becomes classical under the restriction of coarse-grained measurements.
At all times \textit{any} (pure or mixed) spin-$j$ density matrix can be
written in the diagonal form~\cite{Arec1972}%
\begin{equation}
\hat{\rho}=%
{\displaystyle\iint\nolimits_{\Omega}}
\,f(\vartheta,\varphi)\,|\vartheta,\varphi\rangle\langle\vartheta
,\varphi|\,\text{d}^{2}\Omega\label{eq rho}%
\end{equation}
with d$^{2}\Omega\equiv\sin\vartheta\,$d$\vartheta\,$d$\varphi$ the
infinitesimal solid angle element and $f(\vartheta,\varphi)$---usually known
as $P$-function---a \textit{not necessarily positive} real function
(normalization $%
{\textstyle\iint\nolimits_{\Omega}}
\,f(\vartheta,\varphi)\,$d$^{2}\Omega=1$).

The probability for an outcome $m$ in a $\hat{J}_{z}$ measurement in the state
(\ref{eq rho}) is $P(m)=%
{\textstyle\iint\nolimits_{\Omega}}
\,f(\vartheta,\varphi)\,p(m)\,$d$^{2}\Omega$, where $p(m)$ is given by
eq.~(\ref{eq p(m1,t1)}). At the coarse-grained level of classical physics only
the probability for a slot outcome $\bar{m}$ can be measured, i.e., $\bar
{P}(\bar{m})\equiv%
{\textstyle\sum\nolimits_{m\in\{\bar{m}\}}}
P(m)$ with $\{\bar{m}\}$ the set of all $m$ belonging to $\bar{m}$. For
$\Delta m\gg\!\sqrt{j}$ and large $j$ this can be well approximated by
\begin{equation}
\bar{P}(\bar{m})\approx%
{\displaystyle\int\nolimits_{0}^{2\pi}}
\!%
{\displaystyle\int\nolimits_{\vartheta_{1}(\bar{m})}^{\vartheta_{2}(\bar{m})}}
f(\vartheta,\varphi)\sin\vartheta\,\text{d}\vartheta\,\text{d}\varphi
\,,\label{eq P(m) f}%
\end{equation}
where $\vartheta_{1}(\bar{m})$, $\vartheta_{2}(\bar{m})$ are the borders of
the polar angle region corresponding to a projection onto $\bar{m}$. We will
show that $\bar{P}(\bar{m})$ can be obtained from a positive probability
distribution of classical spin vectors. Consider the function%
\begin{equation}
g(\vartheta,\varphi)\equiv\frac{2j+1}{4\,\pi}\,%
{\displaystyle\iint\nolimits_{\Omega^{\prime}}}
\,f(\vartheta^{\prime},\varphi^{\prime})\,\cos^{4j}\!\tfrac{\Theta}%
{2}\,\,\text{d}^{2}\Omega^{\prime}\label{eq g}%
\end{equation}
with d$^{2}\Omega^{\prime}\equiv\sin\vartheta^{\prime}\,$d$\vartheta^{\prime
}\,$d$\varphi^{\prime}$ and $\Theta$ the angle between the directions
$(\vartheta,\varphi)$ and $(\vartheta^{\prime},\varphi^{\prime})$. The
distribution $g(\vartheta,\varphi)$ is \textit{positive} (and normalized)
because it is, up to a normalization factor, the expectation value
Tr$[\hat{\rho}\,|\vartheta,\varphi\rangle\langle\vartheta,\varphi
|]=\langle\vartheta,\varphi|\,\hat{\rho}\,|\vartheta,\varphi\rangle$ of the
state $\left\vert \vartheta,\varphi\right\rangle $. It is usually known as the
$Q$-function~\cite{Agar1981}.

For fuzzy measurements with inaccuracy $\Delta\Theta\sim\vartheta_{2}(\bar
{m})-\vartheta_{1}(\bar{m})\gg1/\!\sqrt{j}$, which is equivalent to $\Delta
m\gg\!\sqrt{j}$, the probability for having an outcome $\bar{m}$ can now be
expressed only in terms of the positive distribution $g$:%
\begin{equation}
\bar{P}(\bar{m})\approx%
{\displaystyle\int\nolimits_{0}^{2\pi}}
\!%
{\displaystyle\int\nolimits_{\vartheta_{1}(\bar{m})}^{\vartheta_{2}(\bar{m})}}
g(\vartheta,\varphi)\sin\vartheta\,\text{d}\vartheta\,\text{d}\varphi\,.
\label{eq P(m) g}%
\end{equation}
The approximate equivalence of eqs.~(\ref{eq P(m) f}) and (\ref{eq P(m) g}) is
shown by substituting eq.~(\ref{eq g}) into (\ref{eq P(m) g}). Note, however,
that $g$ is a mere mathematical tool and not experimentally accessible.
Operationally, because of $\Delta m\gg\!\sqrt{j}$ an averaged version of $g$,
denoted as $h$, is used by the experimenter to describe the system in the
classical limit. Mathematically, this function $h$ is obtained by integrating
$g$ over solid angle elements corresponding to the measurement inaccuracy.
Without the \textquotedblright magnifying glass\textquotedblright\ the regions
given by the experimenter's resolution become \textquotedblright
points\textquotedblright\ on the sphere where $h$ is defined. Thus, a
\textit{full description} is provided by an \textit{ensemble of classical
spins with the probability distribution} $h$.

The \textit{time evolution} of the general state (\ref{eq rho}) is determined
by (\ref{eq Hamiltonian}). In the classical limit it can be described by an
ensemble of classical spins characterized by the initial distribution $g$
($h$), where each spin is rotating according to the Hamilton function
(\ref{eq Hamilton}). From eq.~(\ref{eq P(m) g}) one can see that for the
non-invasiveness at the classical level \textit{it is the change of the }%
$g$\textit{ (}$h$\textit{) distribution which is important and not the change
of the quantum state or equivalently f itself}. In fact, upon a fuzzy $\hat
{J}_{z}$ measurement the state $\hat{\rho}$ \textit{is} reduced to one
particular state, say to $\hat{\rho}_{\bar{m}}$, with the corresponding
(normalized) functions $f_{\bar{m}}$, $g_{\bar{m}}$ and $h_{\bar{m}}$. The
reduction to $\hat{\rho}_{\bar{m}}$ happens with probability $\bar{P}(\bar
{m})$, which is given by eq.~(\ref{eq P(m) f}) or (\ref{eq P(m) g}). Whereas
the $f$-function can change dramatically upon reduction, $g_{\bar{m}}$ is (up
to normalization) approximately the same as $g$ in the region $\Omega_{\bar
{m}}$ between two circles of latitude corresponding to the slot $\bar{m}$ and
zero outside. If $\hat{Q}_{\bar{m}}\equiv%
{\textstyle\sum\nolimits_{m\in\{\bar{m}\}}}
\left\vert m\right\rangle \!\left\langle m\right\vert $ denotes the projector
onto the slot $\bar{m}$, then $\hat{Q}_{\bar{m}}\,|\vartheta,\varphi\rangle$
is \textit{almost} zero ($|\vartheta,\varphi\rangle$) for all coherent states
lying outside (inside) $\Omega_{\bar{m}}$. Thus, $g_{\bar{m}}\propto
\langle\vartheta,\varphi|\,\hat{\rho}_{\bar{m}}\,|\vartheta,\varphi
\rangle\propto\langle\vartheta,\varphi|\,\hat{Q}_{\bar{m}}\,\hat{\rho}%
\,\hat{Q}_{\bar{m}}\,|\vartheta,\varphi\rangle\approx\langle\vartheta
,\varphi|\,\hat{\rho}\,|\vartheta,\varphi\rangle\propto g$ inside and almost
zero outside. Hence, at the coarse-grained level the distribution $g_{\bar{m}%
}$ ($h_{\bar{m}}$) of the reduced state after the measurement can always be
understood approximately as a subensemble of the (classical) distribution $g$
before the measurement. Effectively, the measurement only reveals already
existing properties in the mixture and does not alter the subsequent time
evolution of the individual classical spins.

The disturbance at that level is quantified by how much $g_{\bar{m}}$ differs
from a function which is (up to normalization) $g$ within $\Omega_{\bar{m}}$
and zero outside. One may think of dividing all $g$ distributions into two
extreme classes, i.e., the ones which show narrow pronounced regions of size
comparable to individual coherent states and the ones which change smoothly
over regions larger or comparable to the slot size. The former is highly
disturbed but in an extremely rare fraction of all measurements. The latter is
disturbed in general in a single measurement but to very small extent, as the
weight on the slot borders ($\propto\!\!\sqrt{j}$) is small compared to the
weight well inside the slot ($\propto\!\Delta m$). (In the intermediate cases
one has a trade-off between these two scenarios.) The typical fraction of
these weights is $\!\sqrt{j}/\Delta m\ll1$. Thus, in any case classicality
arises with overwhelming statistical weight.

\textit{Conclusion}.---We showed that the time evolution of an arbitrarily
large spin cannot be understood classically, as long as consecutive outcomes
in a spin component measurement are resolved. For certain Hamiltonians, given
the limitation of coarse-grained measurements, not only is macrorealism valid,
but even the Newtonian time evolution of an ensemble of classical spins
emerges out of a full quantum description of an arbitrary spin state---even
for isolated systems. This suggests that classical physics can be seen as
implied by quantum mechanics under the restriction of fuzzy measurements.

We thank M.~Aspelmeyer, T.~Paterek, M.~Paternostro and A.~Zeilinger for
helpful remarks. This work was supported by the Austrian Science Foundation
FWF, the European Commission, Project QAP (No.\ 015846), and the FWF Doctoral
Program CoQuS. J.~K.~is recipient of a DOC fellowship of the Austrian Academy
of Sciences.


\begin{thebibliography}{99}                                                                                               %


\bibitem {Bell1964}J. S. Bell, Physics (New York) \textbf{1}, 195 (1964).

\bibitem {Legg1985}A. J. Leggett and A. Garg, Phys. Rev. Lett. \textbf{54},
857 (1985); A. J. Leggett, J. Phys.: Cond. Mat. \textbf{14}, R415 (2002).

\bibitem {Schr1935}E. Schrödinger, Die Naturwissenschaften \textbf{48}, 807 (1935).

\bibitem {Zure1991}W. H. Zurek, Phys. Today \textbf{44}, 36 (1991); W. H.
Zurek, Rev. Mod. Phys. \textbf{75}, 715 (2003).

\bibitem {Ghir1986}G. C. Ghirardi, A. Rimini, and T. Weber, Phys. Rev. D
\textbf{34}, 470 (1986); R. Penrose, Gen. Rel. Grav. \textbf{28}, 581 (1996).

\bibitem {Pere1995}A. Peres, \textit{Quantum Theory:~Concepts and Methods}
(Kluwer Academic Publishers, 1995).

\bibitem {Garg1982}A. Garg and N. D. Mermin, Phys. Rev. Lett. \textbf{49}, 901\ (1982).

\bibitem {Yaff1982}L. G. Yaffe, Rev. Mod. Phys. \textbf{54}, 407 (1982); K. G.
Kay, J. Chem. Phys. \textbf{79}, 3026 (1983).

\bibitem {Radc1971}J. M. Radcliffe, J. Phys. A: Gen. Phys. \textbf{4}, 313
(1971); P. W. Atkins and J. C. Dobson, Proc. R. Soc. A \textbf{321}, 321 (1971).

\bibitem {Poul2005}D. Poulin, Phys. Rev. A \textbf{71}, 022102 (2005); P.
Busch, M. Grabowski, and P. J. Lahti, \textit{Operational Quantum Physics}
(Springer, 1995); M. Gell-Mann and J. B. Hartle, Phys. Rev. A \textbf{76},
022104 (2007).

\bibitem {Arec1972}F. T. Arecchi, E. Courtens, R. Gilmore, and H. Thomas,
Phys. Rev. A \textbf{6}, 2211 (1972).

\bibitem {Agar1981}G. S. Agarwal, Phys. Rev. A \textbf{24}, 2889 (1981); G. S.
Agarwal, Phys. Rev. A \textbf{47}, 4608 (1993).
\end{thebibliography}
\end{document}